\documentclass[12pt]{article}

\usepackage[utf8]{inputenc} %% Unicode is now the default for Latex

\usepackage{newtxtext}
\usepackage{amsmath,amssymb,amsthm,cite,enumitem,mathtools,xcolor}
\usepackage{bm} % bold math fonts load last
\usepackage[normalem]{ulem}  %to cross out sth (durchstreichen)

\usepackage[margin=3cm]{geometry}
\newcommand\blfootnote[1]{%
  \begingroup
  \renewcommand\thefootnote{}\footnote{#1}%
  \addtocounter{footnote}{-1}%
  \endgroup}

\numberwithin{equation}{section}

\newtheorem{thm}{Theorem}[section]  %% Theorem 1.1
\newtheorem{prop}[thm]{Proposition} %% Proposition 1.2
\newtheorem{lemma}[thm]{Lemma}       %% Lemma 1.3
\newtheorem{coro}[thm]{Corollary}   %% Corollary 1.4
\newtheorem{remk}[thm]{Remarks}      %% Remark 1.8
\newtheorem{defn}[thm]{Definition}  %% Definition 1.6

\def\eref#1{(\ref{#1})}         
\def\sref#1{Sect.~\ref{#1}}
\def\aref#1{App.~\ref{#1}}
\def\pref#1{Prop.~\ref{#1}}
\def\lref#1{Lemma~\ref{#1}}
\def\rref#1{Remk.~\ref{#1}}

\def\cref#1{Cor.~\ref{#1}}
\def\dref#1{Def.~\ref{#1}}

\def\pa{\partial}

\def\SS{\mathfrak{S}}

\def\sfrac#1#2{\hbox{\large{$\frac{#1}{#2}$}}}

\def\bea#1{\begin{eqnarray}\label{#1}}
\def\eea{\end{eqnarray}}
\def\eref#1{(\ref{#1})}
\def\sref#1{Sect.~\ref{#1}}
\def\ul{\underline}

\def\wh{\widehat}
\def \erw#1{{\langle #1\rangle}}

\def\lra{\leftrightarrow}
\def\ddelta{\boldsymbol{\delta}}
\newcommand{\wick}[1]{{:}#1{:}}
\def\today{}

\title{On the effect of derivative interactions \\ in quantum field theory}

\author{Karl-Henning Rehren$^{1}\quad$% 
\blfootnote{Email: krehren@uni-goettingen.de}
\blfootnote{ORCID: https://orcid.org/0000-0003-3640-6515}
\\[6pt] 
{\footnotesize $^1$Institut für Theoretische Physik, 
Georg-August-Universität Göttingen, 37077 Göttingen, Germany.}
}

\date{\today}

\begin{document}

\maketitle

\begin{abstract}
There exist several good reasons why one may wish to add a total derivative
to an interaction in quantum field theory, e.g., in order to improve the perturbative
construction. Unlike in
classical field theory, adding derivatives in general changes the
theory. The analysis whether and how this can be prevented,
is presently limited to perturbative orders $g^n$, $n\leq 3$. We drastically
simplify it by an all-orders formula, which also allows to answer some salient
structural questions. The method is part of a larger program to (re)derive
interactions of particles by quantum consistency
conditions, rather than a classical principle of gauge invariance. 
\end{abstract}

%\tableofcontents

\section{Introduction}
\label{s:Intro}

\subsection{Motivation} Adding a total
four-derivative $\pa_\mu V^\mu$ (of sufficently rapid decay) to the Lagrangian
density $L(x)$ of a classical field theory does not affect the
Euler-Lagrange equations of motion. The reason is basically that the equations of motion are
equivalent to Hamilton's principle extremalizing the action $\int
d^4x\, L(x)$. The total derivative contributes a boundary term to the
action, that vanishes if $V^\mu$ has sufficiently fast
decay.

The same is not true in quantum field theory. Most notably, the
S-matrix
\bea{STeiL} S_{L_{\rm int}}=Te^{i\int d^4x\, L_{\rm int}(x)}\eea
is sensitive to derivative terms because the time-ordering does not
commute with the time-derivative.

The issue arises, e.g., in the context of BRST theory: The cubic
interaction density $L^{\rm  BRST}_{\rm int}$ 
(including gauge-fixing and ghost terms) in the Standard Model of
particle physics (SM) is BRST
invariant only up to a total derivative.
Therefore, the integral $\int d^4x\, L^{\rm BRST}_{\rm int}$ is BRST invariant, but
it is not obvious that the same is true for its time-ordered
exponential \eref{STeiL}. In this context, the problem was studied
until second order by the Scharf group
\cite{Scha} under the label ``perturbative gauge invariance
(PGI)'': it was discovered that BRST invariance of the
S-matrix requires higher-order interactions (like the quartic
self-coupling of gluons or the self-coupling of the Higgs \cite{GRV}) that can be recursively
determined from the cubic interaction. In this way, parts of the SM were
re-derived without assuming gauge invariance.

Recall that local and covariant massless vector gauge potentials are only defined on
an indefinite state space (Krein space), while local interactions involving massive
vector bosons are non-renormalizable. Thus, in local QFT one is forced to give up
one or the other salient property of quantum field theory:
Hilbert space or renormalizability.
There are then several ways of dealing with massive vector
fields: one may replace them by massless gauge fields and invoke the ``Higgs
mechanism to make them massive''; alternatively, one may replace them
by massive gauge fields plus a scalar Stückelberg field of the same
mass. In both cases one buys renormalizability 
by indefinite metric and extra degrees of freedom, which then have to be
eliminated by BRST. A third way will be addressed next.

The present work is rather motivated by the many recent successes of
``string-localized QFT'' (sQFT) which is a conceptually complementary approach to
BRST. We regard it as ``autonomous'' \cite{Aut} because it is intrinsically quantum,
referring neither to canonical quantization nor to gauge or BRST invariance.
Instead, it is perturbatively defined on the Hilbert space
of the (free) physical particles from the outset. 

\paragraph{String-localized quantum fields.} Rather little of the
technical details of sQFT is actually needed in this paper, but we
want to convey a first idea. It is best illustrated by minimal couplings of conserved 
currents to massless and massive vector bosons:

{\em In QED} -- instead of a canonically quantized local gauge potential
$A_\mu(x)$ that creates unphysical photon states -- one can construct
a quantum vector potential $A_\mu(x,c)$ directly on the physical
photon Fock space. It is given as an integral over the field strength
$F_{\mu\nu}$, of the form
\bea{AFc}
A_\mu(x,c) :=
\int d^4y\, F_{\mu\nu}(x+y) c^\nu(y)
\eea
with  a function $c^\nu(y)$ supported in a ``string'' (a conical spacetime region
emanating from $0$ to spacelike infinity), satisfying $\pa_\nu c^\nu(y)=\delta(y)$. 
Then $\pa_\mu A_\nu(x,c)-\pa_\nu A_\mu(x,c) =
F_{\mu\nu}(x)$ is string-independent.

We 
refer to integrations like \eref{AFc} as ``string integrations''. Such functions $c^\nu$ are not
unique, and varying $c^\nu$, it holds that
$$\delta_c A_\mu(x,c) = \pa_\mu w(x,\delta c)$$
is a derivative. Consequently, the string variation of the minimal
coupling is a total derivative:
\bea{QEDLQ} \delta_c (A_\mu(c)j^\mu) = \pa_\mu(w j^\mu).
\eea
Naively, one would expect that this is sufficient for the S-matrix
\eref{STeiL} to be string-independent:
$\delta_c(S)=0$, as it should (because the function $c^\mu$ is only
auxiliary); but because time-ordering does not commute with
derivatives, this conclusion requires a careful analysis
\cite{Infra}. 

{\em For massive vector particles}, the Proca field $B_\mu(x)$ leads to a
non-renormalizable minimal coupling $B_\mu j^\mu$, whereas a local ``massive
gauge field'' is not defined on a Hilbert space.
Instead, one can construct a string-localized massive vector potential
$A_\mu(x,c)$ on the Hilbert space of the Proca field (by the same string 
integration \eref{AFc} over $F_{\mu\nu}=\pa_\mu B_\nu - \pa_\nu B_\mu$). Its
weaker localization improves its short-distance behaviour, so
that $A_\mu(c)j^\mu$ is power-counting renormalizable. This field
differs from $B_\mu$ by the derivative of another string-localized
field $\phi(c)$:
\bea{ABphi}
A_\mu(x,c)=B_\mu(x) + \pa_\mu \phi(x,c), \quad \phi(x,c):= \int dy\, B_\mu(x+y)c^\mu(y).\eea
Consequently, the minimal couplings differ by a total derivative:
\bea{ProcaLV} A_\mu(c)j^\mu = B_\mu j^\mu + \pa_\mu(\phi(c) j^\mu).
\eea
Again, establishing that the S-matrix is insensitive to the
derivative term, requires a nontrivial analysis.

sQFT can also be applied to QCD \cite{nab} and to perturbative graviton
couplings \cite{GGR}.

sQFT is reviewed in more detail in \cite{Aut}. All that needs to
interest us here is the remarkable fact that all interactions of the SM
can be cast into (generalizations of) one of the above forms, with
which Hilbert space and renormalizability are secured from the
outset. What is more: all such interactions with the 
particle content of the SM involve the familiar cubic interactions of the
SM, e.g., the cubic non-abelian Yang-Mills interactions or minimal
interactions of fermions with vector bosons. 

In this situation, the challenge is to secure string-independence of
the S-matrix, that is: the S-matrix must be insensitive to the
derivatives in interactions like \eref{QEDLQ} or \eref{ProcaLV}. It
turns out that this requires to add higher-order interactions, which
are determined by this condition \cite{GMV,GRV}. This can be worked
out by adapting techniques from the PGI analysis. In all 
cases considered, the higher interactions are again those of the SM,
obtained without invoking a ``gauge principle''.

Because they are necessary to secure
  the string-independence of the S-matrix while preserving the quantum
principles guaranteed by the use of string-localized fields, one may
regard the  higher-order interactions as ``quantum corrections''
accompanying the  derivative terms.

\paragraph{$L$-$Q$ and $L$-$V$ pairs.}

Our intention is to considerably extend the scope, by abstracting from
specific models, and from specific approaches like PGI or sQFT. We consider
two scenarios in which S-matrices must be shown to be insensitive to
derivative terms, called ``LQ'' and ``LV'', respectively,
generalizing \eref{QEDLQ} and \eref{ProcaLV}.
Both scenarios are perturbative in a coupling constant $g$, and the
interaction densities are of the form
\bea{Lint}L_{\rm int}=gL_1+\sfrac{g^2}2 L_2+ \dots,\eea
where $L_n$ are Wick polynomials in free fields. The free fields are
assumed to be given -- either by canonical but possibly indefinite
quantization, or in sQFT constructed by second quantization of the
unitary Wigner representations of the Poincar\'e group, for which a ``free Lagrangian'' plays no role.

{\em In the LQ case,} the S-matrix \eref{STeiL} is required to be invariant
under a derivation $\ddelta$ on the algebra of Wick polynomials:
\bea{SILQ} \ddelta S_{L_{\rm int}}\stackrel !=0. \eea
$\ddelta$ could be the BRST variation $\delta_{\rm BRST}(X)= [Q_{\rm
  BRST},X]_\pm$ as in PGI, or the string-variation $\delta_c$ as in
sQFT. At first order in the coupling constant $g$, this requires $g \int
d^4x\, \ddelta L_1(x) =0$, hence $\ddelta L_1$ must be the
derivative of some quantity $Q_1$ of sufficiently rapid decay:
\bea{LQpair}\ddelta L_1 = \pa_\mu Q_1^\mu.\eea
We call such a structure an ``$L$-$Q$ pair''. Besides the $L$-$Q$
pairs appearing in PGI, the prototype of an $L$-$Q$ pair is that of
string-localized QED, given in \eref{QEDLQ}.
Yet another instance occurs in the QED coupling using Weinberg's
non-covariant and nonlocal vector potential $A_\mu^W(x)$ on the physical photon
Fock space \cite[Sect.~5.9]{Wein} that Lorentz-transforms as
$U(\Lambda)A_\mu^W(x)U(\Lambda)^* = (A_\nu^W(\Lambda x)+ \pa_\nu
\Omega(x,\Lambda)) \Lambda^\nu{}_\mu$, where the operator $\Omega$ is
an unavoidable but physically uninteresting artefact of the construction. Therefore, the interaction
density is not a scalar, because under infinitesimal Lorentz
transformations one has
$\delta_\Lambda(A_\mu^Wj^\mu)= \pa_\mu(\delta_\Lambda(\Omega)j^\mu)$,
and the Lorentz invariance of the S-matrix is at stake \cite{Rivat}
and has to be secured by demanding $\delta_\Lambda(S))=0$. 

Given an $L$-$Q$-pair as a necessary first-order condition, the aim is then to
develop the recursive scheme to determine the ``induced'' interactions
$L_n$ ($n=2,3,\dots$) that are necessary to secure \eref{SILQ}.

{\em The LV case} is more ambitious: By relating two interactions $L_{\rm int}$
and $K_{\rm int}$ (another power series like
\eref{Lint}), it allows to compare and establish equivalences
between two different approaches,
provided the fields of both $L_{\rm int}$ and $K_{\rm int}$ can be
defined on a common space. E.g., sQFT can be compared with gauge
theory by embedding the physical Hilbert space into the Krein
space of gauge theory \cite{Infra}.

Then we are asking for conditions that
\bea{SILV}S_{L_{\rm int}}\stackrel !=S_{K_{\rm int}}.
\eea
Because the first perturbative orders 
$ig\int d^4x\, L_1(x)$ resp.\ $ig\int d^4x\, K_1(x)$ of the
S-matrices in \eref{SILV} do not involve
time-ordering, $L_1$ and $K_1$ can only differ by a total derivative of some quantity $V_1$ of sufficiently rapid decay:
\bea{LVpair}L_1(x) = K_1(x) + \pa_\mu V^\mu_1(x)
\eea
We call such a structure an ``$L$-$V$ pair''. The prototype of an
$L$-$V$ pair in sQFT was given in \eref{ProcaLV}.

Given an $L$-$V$ pair, the aim is to
develop the recursive scheme to determine $L_n$ and $K_n$ ($n=2,3,\dots$) so
that (a strengthened version of) \eref{SILV} is fulfilled: 
$L_{\rm int}$ and $K_{\rm int}$ yield the
same S-matrix. This is particularly interesting when either S-matrix 
manifestly enjoys another 
salient property (like renormalizability and string-independence,
respectively; or Hilbert space and string-independence). If
\eref{SILV} holds,
the S-matrix enjoys both properties.

Every $L$-$V$ pair with $\ddelta (K_1)=0$ (e.g., in sQFT, when $K_1$
is a string-independent local interaction, as in \eref{ProcaLV}) gives
rise to an $L$-$Q$ pair with $Q_1=\ddelta(V_1)$, but the converse is
not true in general.

The LV scenario is much more powerful than LQ
because it allows to establish the equivalence of two different approaches,
rather than just an invariance property of the S-matrix.

A second benefit of the LV scenario will be outlined in \sref{s:main}
and addressed in \sref{s:dress}:
it allows to control the localization of interacting fields in sQFT
which is {\em a priori} at stake because of the nonlocal interaction.

\paragraph{Obstructions and induced interactions.}

The initial $L$-$Q$ pair or $L$-$V$ pair conditions \eref{LQpair} or
\eref{LVpair} arise as necessary conditions for \eref{SILQ} resp.\
\eref{SILV}. They constrain the choice of first-order
interactions (in the coupling constant $g$), given as Wick polynomials
in the free fields. In all
cases of interest, they comprise the cubic part of the interaction. While
this latter feature is irrelevant for the subsequent model-independent analysis, we shall
assume it for definiteness. As a consequence, we shall see that $n$-th
order interactions are Wick polynomials of degree $n+2$.

The pairs \eref{LQpair} resp.\ \eref{LVpair} are the only
input of a model. The common theme in both cases is that higher-order
interactions are then recursively determined (``induced'')
by imposing the validity of \eref{SILQ} resp.\ \eref{SILV} at all orders.
The recursion proceeds in two steps at every order $n$ in the coupling
constant, which we schematically describe for LQ:

Computing the
contributions from all interactions $L_k$ of order $k<n$ to the S-matrix
at order $n$, the result will not satisfy \eref{SILQ}
resp.\ \eref{SILV} in general. The failure is called the {\em obstruction} in
$n$-th order (a Wick polynomial with numeric distributions as
coefficients). E.g., the second order obstruction 
at tree-level, involving two cubic interactions and one Wick contraction, is quartic in the free fields.

The obstruction must have a suitable form, so that it can be cancelled by
adding an interaction term $L_n$ in \eref{Lint}, which can be read off the
obstruction. In this case, we say that the obstruction is ``resolvable'', 
and is resolved by the ``induced'' interaction $L_n$. Otherwise the
condition \eref{SILQ} cannot be fulfilled, and the model has to be
abandoned. The only way to save it is to modify the original $L$-$Q$
pair, which in many cases of interest can be done by allowing further
first-order interactions possibly involving further particles. E.g., the
minimal couplings of non-abelian conserved currents to massive vector
bosons produce non-resolvable obstructions at second order, demanding
an additional $L$-$Q$ pair of cubic self couplings, and resolvability at third order
requires also a cubic coupling to a scalar particle (the Higgs boson). Quartic
self- and Higgs couplings are induced at second order \cite{GRV}.
Moreover, consistency at each order determines some numerical
parameters, including chirality of the weak interaction and
the precise shape of the Higgs self-coupling, see \cite{GMV,GRV} for more details.

The resulting combinations of cubic and quartic interactions are
therefore a consequence of the condition
\eref{SILQ}. Higher-than-quartic interactions (order $n\geq 3$) are
not induced, in accord with the power-counting bound for renormalizability.

Whether the obstruction at any given order is resolvable, is a feature
of the model and has to be decided case by case. All the
interactions of the SM pass this check, and the SM
interactions are to a very large extent determined in this way. There is good
reason to ask why this is so.

We do not know the answer, and we will not attempt to find it in this
work. Instead, we turn to the problem to find a general recursive
formula for the obstruction at 
each order (before it can be evaluated and checked for resolvability), when all lower-order obstructions have been resolved. This
task turned out to be the most difficult step in the recursion, at
least in LV. A first ``pedestrian'' attempt in
\cite{AHM} turned out to be unpracticable beyond the third order. 

We shall give closed formulas at all orders (\pref{p:LQ} for LQ
and \cref{c:LV} and \rref{r:R}.(ii) for LV). This is made possible by
suitable reformulations of the problem, see below. With these
formulas, it is comparatively straight-forward to actually evaluate
the obstructions at tree-level in terms of propagators (time-ordered
two-point functions), and check whether they are resolvable.

Thus, the whole business is about identifying admissible first-order
interactions, {\em and} about determining higher-order from initial
first-order interactions. Recall that the underlying conditions
\eref{SILQ} and \eref{SILV} reflect fundamental principles, notably
Hilbert space via BRST in PGI, and string-independence in sQFT where
the Hilbert space is manifest. Substantial parts of the SM of particle
physics can be built up in this way, and the expectation is that this
is true for the entire SM.

\paragraph{Issues not addressed.}

Out of the scope of the present work are subtleties of analytic
nature, that may arise especially in the actual evaluation of obstructions in sQFT,
based on propagators of string-localized fields. Yet in all case studies based on subtheories
of the Standard Model, these subtleties can be dealt with at least
pragmatically, taking advantage of some freedom of renormalization of
propagators for derivative fields. For a preliminary discussion, see \cite{nab}. A
deeper functional-analytic investigation is certainly needed.

We do also not address loop corrections and UV renormalization,
because the method to determine induced interactions proceeds already
at tree-level. (We have reason to believe that \pref{p:LQ} and
\pref{p:LKU}, relying on the Master Ward Identity (MWI) in \lref{l:MWI},
can be established also at unrenormalized loop level.) The higher-order interactions
thus found are then the starting point for a full renormalized loop analysis, as in
all other approaches. The distinction is that sQFT allows to work with
power-counting renormalizable Hilbert space interactions, even for
couplings of massive vector bosons where local QFT is 
either non-renormalizable, or has to evade into Krein space. 

Finally, we shall not address Chern-Simon’s type boundary terms
  involving fields of slow decay. They are excluded by the assumption
  that the relevant fields have sufficiently fast decay towards
  infinity, so that integrals over total derivatives are zero.

\subsection{A first glimpse at the recursive structure of the problem}
\label{s:glimpse}

\paragraph{LQ setting.} Let an $L$-$Q$ pair \eref{LQpair} be given. The perturbative expansion
of the S-matrix up to second order is
$$S_{L_{\rm int}}= 1 + ig\cdot  \int dx\, L_1(x) + \sfrac {(ig)^2}2\cdot \Big(\iint dx\, dx'\,
T[L_1(x)L_1(x')] -i \int dx\, L_2(x)\Big) + O(g^3).$$
When  the derivation $\ddelta$ is applied, the first-order integral 
vanishes by the $L$-$Q$ pair condition \eref{LQpair}, while the second-order term becomes
\bea{S2LQ}\iint dx\, dx'\,
\big(T[\pa^x_\mu Q_1^\mu(x)L_1(x')]+(x\leftrightarrow x')\big) - i \int dx\, \ddelta
L_2(x).
\eea
The double integral is the contribution from $L_1$ only. It
does not vanish in general because the integrand is not a derivative
because time-ordering
does not commute with derivatives. The question arises whether a suitable
interaction $L_2$ exists whose variation can cancel the double integral
in \eref{S2LQ}.

In order to cancel a double integral by a single integral, one needs
a delta function. It is therefore convenient to subtract from the
former integrand   
the derivative term $\pa^x_\mu T[Q_1^\mu(x)L_1(x')]$. The resulting
\bea{S2O2} O^{(2)}_{\rm LQ}(x,x'):=
\big(T[\pa^x_\mu Q_1^\mu(x)L_1(x')]- \pa^x_\mu T[ Q_1^\mu(x)L_1(x')]\big)+(x\leftrightarrow x')
\eea
is called the ``second-order obstruction of the S-matrix''. The
subtracted term is itself a derivative and does not contribute to \eref{S2LQ}. Yet,
the subtraction is crucial because combinations of the form
\bea{OYX}
O_{Y(x)}(X(x')) := T[\pa^x_\mu Y^\mu(x) X(x')]-\pa^x_\mu T[Y^\mu(x)
X(x')] \equiv [T,\pa_\mu]Y^\mu X' 
\eea
are, for local fields $Y^\mu$ and $X$, supported at $x=x'$, i.e., they exhibit factors $\delta(x-x')$ or derivatives
thereof. As an example, consider the free scalar
  field $\varphi$. Its propagator is the Feynman propagator
  $ i\erw{T[\varphi(x)\varphi(x')]}=\Delta^F(x-x')$, and 
  $i\erw{T[\pa^\mu\varphi(x)\varphi(x')]}=
  \pa^\mu\Delta^F(x-x')$. Compute
  $T[\pa_\mu\pa^\mu\varphi(x)\varphi(x')]-\pa_\mu
  T[\pa^\mu\varphi(x)\varphi(x')]$.
The leading terms according to Wick's theorem $\wick{\pa_\mu
  \pa^\mu\varphi(x) \varphi(x')}-\pa^x_\mu \wick{\pa^\mu\varphi(x) \varphi(x')}$ cancel
out, and because $\square\varphi=-m^2\varphi$, the difference of the contracted terms
is  $$\erw{T[\pa_\mu
  \pa^\mu\varphi(x) \varphi(x')]}-\pa^x_\mu \erw{T[\pa^\mu\varphi(x) \varphi(x')]}
 = i(m^2+\square)\Delta^F(x-x')=i\delta(x-x').$$ 
Expressions of the form \eref{OYX}
frequently appear in QFT. E.g., when $Y^\mu=j^\mu$
is the conserved Dirac current, the 
vanishing of $O_j(X')$ for neutral fields $X$ is well-known as a ``Ward identity'',
while $O_{j(x)}(\psi(x')) =\psi(x)\delta(x-x')$. In sQFT, due to the string-integrations involved in the free fields,
$O_Y(X')$ can be supported on $x'$ 
lying on the string or cone emanating from $x$, or vice versa, see \cite{Aut,nab}.

\begin{defn}\label{d:OYX}

  For $Y$ fixed, we call $X(x')\mapsto O_{Y(x)}(X(x'))$ an
``obstruction map'' on the algebra of Wick polynomials. 
We shall henceforth ony consider the tree-level
contribution, which we shall denote by the same symbol.
\end{defn}
By Wick's theorem, the tree-level contributions
have exactly one contraction. This entails that obstruction maps are
derivations w.r.t.\ the Wick product:
\bea{wickderiv} O_Y(\wick{ X_1X_2}) =\wick{O_Y(X_1)X_2 + 
X_1O_Y(X_2)}
\eea
(facilitating their evaluation on Wick products), and that they
lower the total degree of homogeneity in the free fields by 2.
In particular, if $Q_1$ and $L_1$ are both cubic, then
$O_{Q_1}(L_1')$ is quartic. 

Specifically, turning back to \eref{S2O2}, the second-order
obstruction of the S-matrix is
\bea{O2defLQ}
O^{(2)}_{\rm LQ}(x,x') = O_{Q_1(x)}(L_1(x')) +O_{Q_1(x')}(L_1(x)).
\eea
In order for \eref{S2LQ} to vanish, one must have 
\bea{O2resLQ}
-iO^{(2)}_{\rm LQ}(x,x') \stackrel!= \ddelta L_2(x)\cdot \delta(x-x') - \SS_2\pa^x_\mu
Q^\mu_2(x;x'),
\eea
where $\SS_2f(x,x') = \frac12(f(x,x')+ f(x',x))$ is the
symmetrization in two variables.
$L_2$ and $Q_2$, if they exist, are determined by this cancellation condition
(``resolution of the obstruction''). 

The derivative term $\pa Q_2$ does not contribute to \eref{S2LQ}, but
it is important to keep track of it because the resolution \eref{O2resLQ} will be
used at higher orders of the recursion ``under the $T$-product'',
where derivatives cannot be ignored. This can be seen in the third-order analysis:

The third-order contributions from $L_1$ and $L_2$ to $\ddelta S_{L_{\rm
      int}}$ are given by $\frac{(ig)^3}6$ times
  $$ \iiint\Big(3 T[\ddelta (L_1)L'_1L''_1]
  -3i \delta_{xx'}\big( T[\ddelta (L_1)L''_2] +T[\ddelta(L_2)L''_1]\big)\Big)$$
  with some dummy delta functions $\delta_{xx'}\equiv\delta(x-x')$
  inserted in order to write them as a triple integral.
With the intention to express them in terms of obstruction maps, we subtract
$$0=\iiint \Big(3\pa_\mu T[ Q_1^\mu L_1'L_1''] - 3i\big(\delta_{xx'}\pa_\mu
T[ Q_1^\mu L_2'']+\pa_\mu T[ Q_2^\mu(x;x') L_1'']\big)\Big),$$
and use $\ddelta L_1=\pa Q_1$ and $\delta_{xx'} \ddelta L_2=\SS_2\pa
Q_2-i O^{(2)}$. This yields, in the condensed notation of \eref{OYX}:
$$\iiint 3\Big([T,\pa_\mu]Q_1^\mu L_1'L_1'' -i \delta_{xx'}
[T,\pa_\mu]Q_1^\mu L_2'' -
i[T,\pa_\mu]Q_2^\mu(x;x') L_1'' - T[ O^{(2)}(x,x')L_1''] \Big).$$
By the Master Ward Identity, \lref{l:MWI}, the first term is
$$[T,\pa_\mu]Q_1^\mu L_1'L_1'' =
T[O_{Q_1}(L_1')L_1''] + T[O_{Q_1}(L_1'')L'_1].$$
This equals $T[ O^{(2)}(x,x')L_1'']$ after symmetrization. The terms $T[O^{(2)}L''_1]$ drop out, and we find
the third-order
obstruction expressed entirely in terms of obstruction maps:
\bea{O3defLQ}
O^{(3)}_{\rm LQ}(x,x',x'') = -3i\SS_3 \big(O_{Q_1(x)}(L_2(x'))\cdot \delta_{x'x''} +O_{Q_2(x;x')}(L_1(x''))\big),
\eea
where $O_{Q_2(x;x')}$ is defined as in \eref{OYX} with the derivative $\pa^x$
acting only on $x$. This must be cancelled by the contribution from
$L_3$ up to another derivative, that is:
\bea{O3resLQ}-i^2 O^{(3)}_{\rm LQ}(x,x',x'') \stackrel!= \ddelta L_3(x)\cdot \delta_{xx'x''}  - \SS_3\pa^x_\mu
Q^\mu_3(x;x',x'').
\eea
If this is possible, $L_3$ and $Q_3$ are determined by \eref{O3resLQ}.

We emphasize the importance of the subtraction of derivative terms in
the above, in order to produce the desired forms \eref{O2defLQ} and \eref{O3defLQ}, which
are crucial that there may exist resolutions \eref{O2resLQ} and \eref{O3resLQ}.

\paragraph{LV setting.} The combinatorics in the LV setting is more involved. In
fact, we shall require a strengthened version of \eref{SILV} including
an arbitrary cutoff function, replacing $g$ by $g\chi(x)$ in every
integral. This is necessary because we consider the ``local
S-matrices'' $S[\chi]$ at each order as operator-valued distributions,
to be used for the process of causal renormalization as in the
Epstein-Glaser framework \cite{EG}.

The time-ordered exponential in \eref{STeiL} is a formal expression
that needs UV renormalizations and IR regularizations. The
Epstein-Glaser method regards local S-matrices $S[\chi]$ as
series of symmetric distributions with tensor products of $\chi$ as test
functions. Each $\chi$ is an IR regularization. Products of propagators in loop
contributions are {\em a priori} only defined at non-coinciding points. 
The
UV renormalization can be reformulated as a problem of extension of
distributions in position space to coinciding points, rather than a
subtraction scheme in momentum space.

In this setting, also the obstructions of the
S-matrix at each order are understood as symmetric 
kernels of operator-valued distributions with argument $\chi^{\otimes n}$, i.e., as the
integrands multiplying $\prod_{i=1}^n \chi(x_i)$. 

This prescription requires to include terms of the symbolic form
``$V\circ\pa[\chi]$'' (see \eref{KLV} below, and \cite{AHM}) in one of the two sides in the desired
identity \eref{SILV}, where the derivative acts on one factor $\chi(x_i)$. Such terms obviously vanish in the ``adiabatic limit''
$\chi(x)\to 1$. Because we want to establish a version of \eref{SILV}
for arbitrary $\chi$, we must keep track of these terms and 
integrate by parts all derivatives onto the fields, where they act
outside the time-ordering.

Like the ``subtraction by hand'' of derivatives made in the LQ
setting, the presence of $V\circ\pa$ ensures that all obstructions of the S-matrix
arrange into combinations of obstruction maps. At first order,
\bea{SILV!}S_{L_{\rm int}+ V\circ\pa}[\chi]-S_{K_{\rm int}}[\chi] = ig \int dx\,
\big((L_1(x)-K_1(x))\chi(x)+V^\mu_1(x)\pa_\mu \chi(x)\big) + O(g^2)
\eea
vanishes because $(L_1-K_1)\chi + V_1\pa\chi = \pa(V_1\chi)$ is a
total derivative, thanks to the $L$-$V$ pair
condition. The expansion of \eref{SILV!} at second order gives
$$\sfrac{(ig)^2}2\iint dx\, dx' \Big(T[(L_1+V_1\pa^x)(L_1'+V_1'\pa^{x'})]- T[K_1K_1'] - i
\big((L_2-K_2)\delta_{xx'}+ V_2\pa^x\big)\Big)\chi(x)\chi(x').$$
The derivatives act on the test functions. 
Integrating them by parts (so
that they act outside the $T$-product) and using $L_1=K_1+\pa V_1$, one
gets 
\bea{S2LV}
&& S_{L_{\rm int}+ V\circ\pa}[\chi]-S_{K_{\rm int}}[\chi] \\ \notag && = 
\sfrac{(ig)^2}2 \cdot \iint dx\, dx'\,
\Big(O^{(2)}(x,x') - i\big((L_2(x)-K_2(x))\delta_{xx'}-\pa_\mu^x V^\mu_2(x;x')\big) \Big)\chi(x)\chi(x') + O(g^3),
\eea
where $O^{(2)}(x,x') $ is the distributional kernel of the
  second-order contributions from $L_1$, $K_1$
and $V_1$:
\bea{O2defLV}
O^{(2)}(x,x') = \SS_2 \big(O_{V_1}(L'_1+K'_1) - \pa^x_\mu
O_{V'_1}(V^\mu_1)\big).
\eea
This quantity is called the second-order obstruction. It must be cancelled by
\bea{O2resLV} -iO^{(2)}(x,x') \stackrel!= (L_2(x)-K_2(x))\cdot \delta_{xx'}  - \SS_2\pa^x_\mu
V^\mu_2(x;x').\eea
$L_2$, $K_2$ and $V_2$, if they exist with the respective specified 
``salient properties'' as explained in the motivations, are
determined by this resolution of the obstruction.

The expression for the third-order obstruction $O^{(3)}(x,x',x'')$ in
terms of obstruction maps is already quite intricate. It  was computed
in \cite{AHM} (again with the use of the MWI, and recursive use of the
second-order resolution) under model-specific assumptions on the form of
$V_2^\mu$, and found to involve iterated obstruction maps 
$O_{V_1}\circ O_{V_1'}(\cdot)$ as well as ``nested'' obstruction maps
$O_{O_{V_1}(V_1')}(\cdot)$.

\subsection{Main results}
\label{s:main}

\paragraph{All-order obstruction formulas.} The  primary aim of this
paper is to give expressions for the obstructions of the S-matrix at 
all orders in terms of obstruction maps involving lower-order
  interactions, generalizing \eref{O2defLQ}, \eref{O3defLQ}, and
\eref{O2defLV}. This task is comparatively easy in the LQ setting 
\cite{Mü}.  In the LV setting, it is made possible by a drastic
simplification (as compared to the first attempt in \cite{AHM}) due to an advantageous 
``reparametrization'' that takes substantial {\em a priori} cancellations into account. To illustrate the gist at second order: the
manifest derivative term in the obstruction \eref{O2defLV} would be part of
the derivative term $\pa V_2$ in \eref{O2resLV} that must resolve the
obstruction. Equivalently, the obstruction is resolved by finding
$L_2$, $K_2$ and $U_2$ that solve the simpler equation
$$-i\SS_2 O_{V_1}(L'_1+K'_1) \stackrel!=(L_2-K_2)\cdot
\delta_{xx'} - \SS_2\pa^x_\mu U_2^\mu(x;x'),$$
where $U^\mu_2(x;x') = V^\mu_2(x;x') +i O_{V'_1}(V_1^\mu)$ is 
  typically a simpler expression  than $V_2$. The ``nested''
obstruction map $O_{O_{V_1}(V_1')}$ appearing at 
third order will be absorbed in $O_{U_2}$. The point is that the same will happen
recursively at all higher orders. Therefore, starting from an $L$-$V$ pair involving
  $U_1:=V_1$, we arrive at a far more transparent recursion determining $L_n$, $K_n$ and $U_n$.

Apart from this simplification, the present setup admits the
derivative terms in the resolutions to be fields of
the form $Q^\mu_n(x;x_2,\dots,x_n)$ in LQ and $V^\mu_n(x;x_2,\dots,x_n)$
(or $U_n$) in LV
that are supported at $x_i=x$ or (in the sQFT context) at $x_i$ on the
string emanating from $x$.
It was observed in models with non-abelian self-interactions of vector
bosons that such fields necessarily appear. This generalization does not
harm because $Q_n$ do not appear in the S-matrix, and the 
contributions from $U_n$ will disappear in the adiabatic limit.

\paragraph{The dressed field.} In the LV scenario, two different
interactions are compared that give rise to the same S-matrix. Yet,
the interacting fields constructed perturbatively (by ``Bogoliubov's
formula'', see \sref{s:dress}) will not be the same with both interactions.

Instead, we shall construct in \sref{s:dress} a coupling-constant dependent
``dressing transformation'' (in fact, an automorphism) of the Wick algebra of free fields
\bea{dress1}
\Phi(x)\mapsto\Phi_{[g]}(x)
\eea
such that it holds for the respective interacting fields
\bea{magic1}
\Phi\vert_{L_{\rm int}}(x)=\Phi_{[g]}\vert_{K_{\rm int}}(x).\eea
This formula is particular important in sQFT where it addresses a
vital and critical issue: with nonlocal interactions like $L_{\rm int}(c)$, one 
{\em a priori} looses control over the localization of interacting fields
$\Phi\vert_{L_{\rm int}}(x)$. On the other hand, when an $L$-$V$ pair
with a local interaction $K_{\rm int}$ (e.g., in a BRST setting) is given, the dressed field $\Phi_{[g]}$ is at worst 
string-localized. Because $K_{\rm int}$ is local, it preserves
relative localizations of interacting fields.

Thus, one can conclude from \eref{magic1} that $\Phi\vert_{L_{\rm int}}(x)$ are
in general string-localized, while free fields for which $\Phi_{[g]}$ turn out
to be local, give rise to the local interacting observables
$\Phi\vert_{L_{\rm int}}(x)$.

This is a great benefit as compared to BRST, where interacting fields
that are not BRST invariant (like charged fields), are just not defined on the Hilbert
space. In sQFT, they are defined on the Hilbert space but with a
weaker localization. The latter feature is physically meaningful, e.g., in
order to resolve the conflict between Gauß' Law and Locality \cite{FPS}, see the
first example in \aref{a:Exa}.

The existence and relevance of
string-localized charged fields in Nature were anticipated abstractly by the
analyis of the localization structure of charged superselection sectors in Algebraic
QFT \cite{BF}. By \eref{magic1}, sQFT provides a handle to actually construct them
(perturbatively).

\subsection{Outline of the paper}
\label{s:outline}

We write $L$ for $L_{\rm int}$ and $K$ for $K_{\rm int}$ from now on.

Because the recursive determination of higher-order interactions is a
condition for consistency with fundamental principles,
we shall impose the validity of \eref{SILQ} and \eref{SILV} only at
tree level, as a {\em necessary} condition. In particular, we do not
address UV renormalization of loops in this paper. Also, all
obstruction maps are understood at tree level.  

In the LQ scenario (\sref{s:LQall}), the
adiabatic limit $\chi(x)\to 1$ is understood, and we shall suppress
the cutoff function $\chi$ altogether. We find 
the recursive structure of the obstructions $O^{(n)}_{\rm LQ}$ at all orders (\pref{p:LQ}), from
which the interactions satisfying $\ddelta S_L=0$ can be
determined. Here, the property that
$\ddelta$ in \eref{LQpair} is a derivation, is instrumental, as well as the MWI, \lref{l:MWI}.

In the LV scenario (\sref{s:LVall}) we are more ambitious by imposing the validity of 
\eref{SILV} with an arbitrary cutoff function $\chi(x)$, as a prerequisite for the UV
renormalization at loop level. Having to deal with two interactions,
complicates the task further, but the reparametrization $V\to U$
announced in \sref{s:main} helps. More importantly, there is no
derivation $\ddelta$ at our  disposal. While this derivation was
instrumental in LQ, we find a ``substitute'' in the
form of an interpolation \eref{LVt} between $K$ and $L$ with
a parameter $t$, such that the derivation property of $d/dt$ can be
exploited.

\pref{p:LKU} then provides a formula \eref{LKU} for $L$ as a power series in the
obstruction map $O_U$, acting on $K$ and on $\pa_\mu U^\mu$.
This formula is the first main result. In fact, the field $V$ is not needed
for this formula. For the validity of \eref{LVchi}, it suffices to know
that $V$ exists and can be computed as a power series in $U$. Its precise form is
uninteresting because it disappears in the adiabatic limit anyway.

The main formula \eref{LKU} can be rewritten in several
ways. Particularly useful is a rewriting in such a way
(Cor.~3.4) that the
obstructions of the S-matrix at each perturbative order
$g^n$  can be directly read off as iterated obstructions
$\Pi_iO_{U_{k_i}}$ acting on $L_k$ and $K_k$ (with
$k+\sum_ik_i=n$). This surprisingly simple formula at all orders
is the main technical improvement 
over the clutter in \cite{AHM} already at $n=3$.

In \sref{s:dress}, we establish a formula for the
``dressed field'' at all orders, and draw some interesting conclusions
from it. Nontrivial examples can be found in the appendix. 

For relevant applications of the $L$-$Q$ formalism to the bosonic and
fermionic sectors of the weak interactions, see
\cite{GGM,GMV,GRV,AHM}.
For interesting and nontrivial applications of the
$L$-$V$ formula, and of the dressing transformation, see \aref{a:Exa},
notably the third example.

\section{The all-orders $L$-$Q$ formula}
\label{s:LQall}

Working in perturbation theory means working with formal power series
in the coupling constant $g$. The interactions $L=L(g)$ in \eref{SILQ} and
\eref{KLV} are themselves power series (and in fact in most cases
polynomials) in $g$ with leading order $g$. Since power series in
power series in $g$ are again power series in $g$, it 
is legitimate, and will turn out to be most advantageous, 
to work and display the results in terms of products and exponentials
of power series like $L(g)$. The
order-by-order decomposition into orders $g^n$ is then a trivial step.

We shall lift $L_n(x)$ to $\wh
L_n(x_1,\dots,x_n)=L_n(x_1)\delta_{x_1,\cdots,x_n}$, and suppress all arguments that are integrated over with the symbol $\iint$. This is just a device to unburden the notation, enabling us to concentrate on the recursive structure. The
arguments etc.\ can be unequivocally re-installed at the end, with
\eref{O3defLQ} and \eref{O3resLQ} as a ``blueprint'':  each $L_k$ on
the right-hand side comes with an argument $x$ and a total
delta function in $k$ arguments, and each $Q_k$ has one
distinguished argument (on which the derivative acts) and is
symmetric in its remaining $k-1$ arguments. The whole expression is
symmetrized in $n$ arguments.

By the derivation property of $\ddelta$, the desired invariance of the
S-matrix \eref{SILQ} can be written as
\bea{dSdL} \ddelta S_L=i\int dx\, T[\ddelta L(x) \cdot e^{i \int L}]\stackrel!=0. \eea
At order $g^N$, this is
\bea{dSN}\ddelta S_L^{(N)}= 
\sfrac{g^N}{N!}\Big[i\int dx\, \ddelta L_N(x) + i^N\iint\hbox{contributions from lower
  interactions $L_k$ ($k<N$)}\Big].\eea
Recall that when the integral over the ``lower contributions'' does not
vanish, then $\ddelta S_L$ cannot vanish with the lower interactions
alone. But because these contributions alone, being integrals over
propagators, cannot be cancelled by an
integral over $\ddelta L_N(x)$, they must be prepared further to
become a resolvable obstruction, following the strategy in
\sref{s:glimpse}.

We begin with an informal
``definition'' of the $n$-th order LQ obstruction $O^{(N)}_{\rm
  LQ}$. It is clear from the examples $n=2,3$ (see \eref{O2defLQ} and
\eref{O3defLQ} in \sref{s:glimpse}) that
$O^{(N)}_{\rm LQ}$ can only be defined when all lower obstructions
have already been resolved. 
\begin{defn}\label{d:ONLQ} $\frac{(ig)^N}{N!}\cdot O^{(N)}_{\rm LQ}(x_1,\dots,x_N)$
  is the integrand of the contribution from all interactions $L_k$
  ($k < N$) to the $N$-th order of $\ddelta S_{L}(g)$,
  recursively prepared by the subtraction of suitable derivatives of
  terms involving lower interactions $L_k$ and $Q_k$ ($k<N$), in
  such a way that it is expressed in terms of obstruction maps
  $O_{Q_{k_1}}(L'_{k_2})$.
  \end{defn}
That the integrand can be prepared as asserted provided the lower
obstructions have been resolved (as we have
seen at orders $n=2,3$), will become manifest in \pref{p:LQ}.

To find the formula for the yet unknown $O^{(N)}_{\rm LQ}$, let us
fix some $N$ and assume that all $O^{(n)}_{\rm LQ}$ ($n<N$) are known
and all obstructions of order $n<N$ have be resolved:
\bea{LQresn} i^{n}O^{(n)}_{\rm LQ} +i (\ddelta L_n-\pa_\mu Q^\mu_n)
=0.
\eea
We formally write the partial sums
\bea{resum}\ul{\wh L} := \sum_{n=1}^{N-1}  \frac{g^n}{n!}\wh L_n, \qquad \ul{Q^\mu} := \sum_{n=1}^{N-1}  \frac{g^n}{n!}Q_n^\mu, \qquad 
\ul{O_{\rm LQ}} := \sum_{n=2}^{N-1} \frac{i^{n-1}g^n}{n!}O_{\rm
  LQ}^{(n)}.\eea
Thus, \eref{LQresn} for all $n<N$ becomes
\bea{LQresN} \ul{\ddelta \wh L} - \pa_\mu \ul{Q^\mu} +\ul{O_{\rm LQ}} =0.\eea
We now subtract from \eref{dSdL} an integral over a derivative:
\bea{dSsub}
\ddelta S_L=i\iint T[\ddelta \wh L\cdot e^{i \int L}] - i\iint
\pa_\mu T[\ul{Q^\mu} e^{i\int L}],
\eea
which at order $N$ subtracts derivatives of
terms involving lower interactions $L_k$ and $Q_k$ ($k<N$), as
suggested by \dref{d:ONLQ}.

We insert \eref{LQresN} into \eref{dSsub}, exhibiting the explicit
term $\ddelta L_N$ not contained in \eref{LQresN}, and neglecting higher-order terms:
\bea{dSOQ}\ddelta S_L=i\iint \Big(T [-\ul{O_{\rm LQ}}e^{i\int L}] +  T[\pa_\mu \ul{Q^\mu}
e^{i\int L}] - \pa_\mu
T[\ul{Q^\mu} e^{i\int L}]\Big) + \frac{g^N}{N!} i\int \ddelta L_N + O(g^{N+1}).\eea

At this point, we need the MWI, which holds identically at tree level
by virtue of Wick's theorem. In fact, it also holds at unrenormalized
loop level and is often imposed as a renormalization condition
\cite{Dü} in local QFT:

\begin{lemma}\label{l:MWI} (Master Ward Identity, see \cite{Dü}). For Wick
polynomials $Y$ and $X_i$, it holds
\bea{MWI} \notag 
O_{Y}(X_1,\dots,X_m)&:=& T[\pa_\mu Y^\mu(x) \cdot X_1(x_1)\ldots
X_m(x_m)]-\pa^x_\mu T[Y^\mu(x) \cdot X_1(x_1)\ldots
X_m(x_m)]
\\  &=& \sum_{i=1}^m T[ X_1(x_1)\ldots
\cdot O_{Y(x)}(X_i(x_i))\cdot \ldots X_m(x_m)].
\qquad\eea
\end{lemma}
In sQFT, the MWI must be extended {\em mutatis mutandis} to string-localized
fields $X(x,c)$, and to multi-local fields
$Y_k^\mu(y;y_2,\dots,y_k)$, such that the derivative is on the first
argument $y$. The former extension is unproblematic when
the time-ordering is w.r.t.\ the ``apex'' $x$ of the string-localized
fields $X(x,c)$. The latter extension has not been established
in full generality, but could be done in all applications of sQFT sofar, see the discussion
in \cite{nab}, and also \cite{Mü}.  We thus assume the MWI to hold
also in sQFT.

The ``derivation-like'' structure of \eref{MWI} implies

\begin{coro}\label{c:TpaYE} It holds
\bea{}
T[\pa_\mu Y^\mu(x) e^{i\int L}] - \pa^x_\mu T[Y^\mu(x) e^{i\int L}] =
i\int dx\,  T [O_{Y}(L(x))
e^{i\int L}].
\eea
\end{coro}
{\em Proof:} Expand the exponential as a power series and apply
\lref{l:MWI} to each term. \qed

Because derivatives do not matter, \cref{c:TpaYE} with $\ul Q$ in the place of $Y$ turns \eref{dSOQ} into
\bea{dSOO}
\ddelta S_L=i\iint T [-\ul{O_{\rm LQ}}e^{i\int L}] -\iint T[O_{\ul
  Q}(\wh L) e^{i\int L}] + i\frac{g^N}{N!} \int \ddelta L_N + O(g^{N+1}),
\eea
where the integrations in the second term are also over the arguments of
$\wh L$.

Now, we can prove
\begin{prop}\label{p:LQ} (All-orders LQ obstruction). If all
  obstructions are resolvable, then
\bea{LQn} O^{(n)}_{\rm LQ} = i^{2-n} \sum_{k_1+k_2=n}
\sfrac{n!}{k_1!k_2!} \, \SS_kO_{Q_{k_1}}(\wh L_{k_2}),
\eea
where obviously all $k_1,k_2<n$. \\
More precisely: $\ddelta S_L$ vanishes, if and only if at each order $L_n$ and
$Q_n$ solve the recursive system \eref{LQresn} with $O^{(n)}_{\rm
  LQ}$ given by \eref{LQn}.
\end{prop}

{\em Proof by induction:} For $n=1$ ($O^{(1)}_{\rm LQ}=0$) and
$n=2,3$, \eref{LQn} was shown in \sref{s:glimpse}, ensuring the
vanishing of $\ddelta S_L$ at these orders.

When \eref{LQn} is assumed to hold for all $n<N$, then $-i\ul{O_{\rm LQ}}$ in \eref{dSOO}
is a sum
$$\sum_{k_1,k_2}
\sfrac{g^{k_1+k_2}}{k_1!k_2!} \, \SS_kO_{Q_{k_1}}(\wh L_{k_2})
$$extending over $k_1+k_2<N$. On the other hand, $O_{\ul
  Q}(\wh L)$ in  \eref{dSOO} is the same sum, but extending over
$k_1<N$ while $k_2$ is arbitrary. Thus, all terms with $k_1+k_2<N$
cancel in \eref{dSOO}, and precisely the terms with $k_1+k_2=N$
survive at order $g^N$. This turns \eref{dSOO} into
\bea{dSfin}
\ddelta S_L=-\iint T[ \sum_{k_1+k_2=N}
\sfrac{g^{N}}{k_1!k_2!} O_{Q_{k_1}}(\wh L_{k_2})e^{i\int L}] + \frac{g^N}{N!} i\int \ddelta L_N + O(g^{N+1}),
\eea
and the factor $e^{i\int L}$ can be omitted at $N$-th order.

Thus, the integrand at $N$-th order has been prepared to become
\bea{dSint}
\frac{g^N}{N!} \big(i^NO^{(N)}_{\rm LQ}(x_1,\dots,x_n) + i\ddelta 
L_N \cdot \delta_{x_1,\dots,x_n}\big)
\eea
with $O^{(N)}_{\rm LQ}$ again given by \eref{LQn}. For $\ddelta S_L$
to vanish at $N$-th order, \eref{dSint} must be a total derivative. This
yields also \eref{LQresn} for $n=N$.

This concludes the
proof by induction. \qed

  \begin{remk} \label{r:pcb} \rm
(i) By inspection of the short-distance scaling dimensions (which behave
additively unter time-ordered products, hence under obstruction maps) of the expressions \eref{LQn}, one can see by induction that if $L_1$ 
satisfies the power-counting bound for renormalizability, so do
all induced $L_n$ -- provided all
obstructions are resolvable. On the other hand, because their
degree as Wick polynomials increases with $n$, induced
interactions $L_n$ must vanish for $n>2$.%
\footnote{\label{fn:pcb} I thank M. Dütsch who pointed out this argument.}
\\[1mm]
This does not mean that \pref{p:LQ} is void for $n>2$. Namely, one
must still evaluate the higher order obstructions in order to show
that they are resolvable with $L_n=0$; i.e., they must be total derivatives.
\\[1mm]
(ii) On the other hand, graviton couplings
\cite{GGR} are an instance with $L_1$ non-renormalizable, and the
series of $L_n$ does not terminate. Indeed, $L_2$ has been computed to
coincide with the cubic expression expected from classical GR, and the
same is expected for $L_n$ (see \cite{Dü2} for a similar result in PGI.)
  \end{remk}

\section{The all-orders $L$-$V$ formula}
\label{s:LVall}

In the presence of the cutoff function, i.e., before the
  adiabatic limit is taken, \eref{SILV} cannot hold. Instead, we shall demand the
equality
\bea{LVchi}
\big(Te^{iK}\big)[\chi] = \big(Te^{i(L+V_\mu\circ\pa^\mu)}\big)[\chi],
\eea
where (in distributional notation)
\bea{KLV}
K[\chi] = \sum_{n\geq 1} \sfrac{g^n}{n!}K_n(\chi^n), \quad
L[\chi] = \sum_{n\geq 1} \sfrac{g^n}{n!}L_n(\chi^n), \quad
V^\mu\circ\pa^\mu[\chi]=\sum_{n\geq 1}
\sfrac{g^n}{n!}V^\mu_n(\pa_\mu\chi\otimes\chi^{\otimes (n-1)}).\quad
\eea
When $V_n$ appears inside a time-ordered product with test
function $\pa\chi\otimes \chi^{\otimes (n-1)}$, 
one can integrate the derivative by parts so that it acts on the time-ordered product of
fields. Then, all test functions are just tensor powers of $\chi$,
which means that the corresponding distributional kernels are
completely symmetric. The symbol $V_\mu\circ\pa^\mu$ is always understood in this sense.

Clearly, in the adiabatic limit $\chi\to1$, \eref{LVchi} becomes \eref{SILV}.
The task is to find the recursive structure for the
  obstructions, that allow to determine  the interactions
  $L$ and $K$ (and $V$) such that \eref{LVchi} can hold.

The more general form of $V_n(x_1;x_2,\dots,x_n)$ as compared to the ansatz in \cite{AHM}
means a gain in flexibility to satisfy \eref{LVchi}, that is
unavoidable in non-abelian models of sQFT \cite{nab}, and at the same time
harmless because $V$ becomes irrelevant in the adiabatic limit.

The problem will be attacked by proving that along with \eref{LVchi} there
holds an interpolation
between $K$ and $L+V\circ\pa$, see \pref{p:LVt}.
In ``solving'' \eref{LVt} for the interpolating interactions $\ell(t)$
and $v(t)$, from which ultimately $L$, $K$ and $V$ can be determined, an invertible substitution of power series $V^\mu\lra U^\mu$
turns out to be very helpful. The passage from $V$ to $U$ is the ``change of variables'' mentioned
in \sref{s:main}. In particular, it will absorb all ``nested'' obstruction maps.

We adopt a similar efficient 
short-hand notation as in \sref{s:LQall}, now keeping track of the
cutoff function. We think of $L_n(\chi^n)$ as $(L_n\delta^{(n)}_{\rm
  tot})(\chi^{\otimes n})$ (and
similarly with $K$), so that all distributions at order $n$ are evaluated on
$\chi^{\otimes n}$. 

\begin{prop}\label{p:LVt} (Interpolation). Suppose that \eref{LVchi} holds:
\bea{LVprop}
\big(Te^{iK}\big)[\chi] = \big(Te^{i(L+V^\mu\circ\pa_\mu)}\big)[\chi]
\eea
with power series (in $g$) of Wick polynomials $L$, $K$, $V^\mu$ as in
\eref{KLV}. 
Then there exist power series of Wick polynomials
$\ell(t)$ and $v^\mu(t)$ depending on a parameter $t\in[0,1]$, such that also
\bea{LVt}
\big(Te^{iK}\big)[\chi] = \big(Te^{i(\ell(t)+v^\mu(t)\circ\pa_\mu)}\big)[\chi]
\eea
holds, satisfying the boundary conditions $\ell(0)=K$, $v^\mu(0)=0$ and
$\ell(1)=L$, $v^\mu(1)=V^\mu$.
\end{prop}
For the proof, we shall need another Lemma, which states the solution
of a simple inhomogeneous linear 
differential equation. It is elementary to prove:

\begin{lemma}\label{l:IVP} Let $M$ be a linear map on some vector space $S$. Then
the inhomogeneous initial value problem for a function $s(t)\in S$
\bea{DE}
\dot s(t) - M(s(t)) = a \quad \hbox{with}\quad s(0)= s_0
\eea
is solved (as a power series in $t$) by
\bea{DEsol}
s(t) = \exp (tM)(s_0) + F(tM)(ta), \quad \hbox{where}\quad F(u):=
\sfrac{e^u-1}u = \sum_{r=0}^\infty \sfrac{u^r}{(r+1)!}.
\eea
\end{lemma}
{\em Proof of \pref{p:LVt}:} For any power series of Wick polynomials $\ell$, $v$, and
$U$, \cref{c:TpaYE} with $L$ replaced by $(\ell+v\circ\pa)[\chi]$ yields
\bea{OUlv}\big([T,\pa_\mu] U^\mu e^{i(\ell+v\circ\pa)}\big)[\chi] =
\big(T[iO_U(\ell)e^{i(\ell+v\circ\pa)}] - \pa_\mu
T[iO_U(v^\mu)e^{i(\ell+v\circ\pa)}]\big)[\chi],\eea
equivalently
$$\big(T\big[\big(\pa_\mu U^\mu -iO_U(\ell)\big)e^{i(\ell+v\pa)}\big]-\pa_\mu T\big[\big(U^\mu-iO_U(v^\mu)\big)e^{i(\ell+v\pa)}\big]\big)[\chi]=0.$$
Consequently, if functions $\ell(t)$ and $v(t)$ satisfy the differential equations
\bea{ldotvdot}
\dot \ell(t) = \pa_\mu U^\mu -iO_U(\ell(t))\quad\hbox{and}\quad
\dot v^\mu(t)= U^\mu-iO_U(v^\mu(t)),
\eea
then \eref{LVt} is constant, by virtue of \eref{OUlv} and \eref{ldotvdot}:
\bea{ddt}\sfrac d{dt} \big(Te^{i(\ell(t)+v^\mu(t)\circ\pa_\mu)}\big)[\chi] =
\big(T[\dot\ell(t)e^{i(\ell(t)+v(t)\pa)}] - \pa_\mu T[\dot
v^\mu(t)e^{i(\ell(t)+v(t)\pa)}]\big)[\chi]=0.\eea

The proof of \pref{p:LVt} therefore amounts to the existence of
$U^\mu$ and $\ell(t)$, $v^\mu(t)$ satisfying \eref{ldotvdot}, such
that the boundary conditions $\ell(0)=K$, $v^\mu(0)=0$ and $\ell(1)=L$, $v^\mu(1)=V^\mu$
are fulfilled.

By virtue of \lref{l:IVP}, the initial values at $t=0$ fix
\bea{ellvt}
\ell(t) =  \exp(-itO_U)(K) + F(-itO_U)(t\pa_\mu U^\mu), \qquad v^\mu(t) = F(-itO_U)(tU^\mu).
\eea
The final values demand
\bea{ellv1}
\ell(1) =  \exp(-iO_U)(K) + F(-iO_U)(\pa_\mu U^\mu)\stackrel!=L, \qquad
v^\mu(1) = F(-iO_U)(U^\mu) \stackrel!= V^\mu.
\eea
The latter equation relates $V$ to $U$:
\bea{VU}
V^\mu = F(-iO_U)(U^\mu) = U^\mu - \sfrac i2 O_U(U^\mu) -\sfrac 16 O_U^2(U^\mu)+ \dots.\eea
This power series can be inverted:
\bea{UV}
U^\mu = V^\mu+\sfrac i2 O_V(V^\mu) -\sfrac1{12}
O_{V}^2(V^\mu)-\sfrac 14 O_{O_V(V)}(V^\mu) + \dots,
\eea
and therefore determines $U$ in terms of $V$. Since $V$ is itself a
power series in the coupling constant $g$, so is $U$. Thus, \eref{ellvt}
establishes \pref{p:LVt}, provided also the first relation in
\eref{ellv1} holds.

For this purpose, it suffices to notice that
when $K$ and $V$ in \eref{LVprop} are given, then $L$ is unique if it
exists. Namely, at each perturbative order the right-hand side of \eref{LVprop} is a sum of $L_n$ and
terms involving only $V_k$, and $L_k$ with $k <n$,
which fixes $L_n$ recursively. Therefore, since \eref{LVt} at $t=1$ is \eref{LVprop} with $\ell(1)$ in the place of $L$, it follows that $\ell(1)=L$.
  \qed

\begin{prop}\label{p:LKU} (``LV identity''). (i) The identity \eref{LVprop}
holds if and only if 
\bea{LKU}
L =\exp( -iO_U)(K) + F(-iO_U)(\pa U),\eea
with $U=U(V)$ given by the inverse of the power series
$V(U)=F(-iO_U)(U)$ as in \eref{VU}, where
$F(u)=\sfrac{e^u-1}u$ as in \lref{l:IVP}. 

(ii) It also holds
\bea{KLU}
K= \exp(iO_U)(L) - F(iO_U)(\pa U),
\eea
exhibiting a perfect symmetry under $L\leftrightarrow K$ and
$U\leftrightarrow -U$.
In particular, \eref{LVprop} is also equivalent to
$$Te^{iL[\chi]}= Te^{i(K- \widetilde V\circ\pa)[\chi]}$$
with $\widetilde V = F(+iO_U)(U)$. 
\end{prop}
{\em Proof:} The first formula \eref{LKU} is just the evaluation of the solution
$\ell(t)$ in \eref{ellvt} at $t=1$. Therefore, \eref{LVprop} implies \eref{LKU} by
\pref{p:LVt}. Conversely, the proof of \pref{p:LVt} shows that \eref{LVt}
with $\ell(t)$ and $v(t)$ given by \eref{ellvt} is independent of
$t$. Putting $t=1$, one gets \eref{LVprop}.

For the second formula \eref{KLU}, we solve \eref{LKU} for $K$: we write $F(u) =
\int_0^1dt \, e^{tu}$ in \eref{LKU}, apply $e^{iO_U}$ on both sides,
and substitute $1-t\to t$ in the integral. \qed

One can also solve \eref{LKU} for $\pa U$
and then rewrite it in a way such that the obstruction maps
act only on the interactions $L$ and $K$:

\begin{coro}\label{c:LV} (All-orders LV obstruction). With $G(u)$ the even power series $G(u)=1-\frac12u  \cot (\frac12
u)= \frac1{12}u^2+ \frac1{720}u^4+\dots$, it holds 
 \bea{LpmK}L-K-\pa_\mu U^\mu = G(O_U)(L-K)
- \sfrac i2O_U(L+K). \eea
\end{coro}
{\em Proof:}
Apply $\exp (\sfrac i2O_U)$ to \eref{LKU}, and solve the resulting
$$  \exp (\sfrac i2O_U)(L) - \exp (-\sfrac i2O_U)(K) = iH(\sfrac12 O_U)(\pa U),
\quad\hbox{where} \quad H(u) := \frac{\sin(u)}u $$
for $\pa U$ by using that the inverse of
a power series of maps $1+\dots$ is the reciprocal power series.
The reason why $L-K$ is added on both
sides of \eref{LpmK} will be clear from \rref{r:R}.(ii). \qed

\begin{remk}\label{r:R}\rm 
(i) \cref{c:LV} manifestly exhibits the symmetry mentioned in \pref{p:LKU}. \\[1mm]
(ii) \cref{c:LV} is used to recursively determine the induced higher-order
interactions. By expanding the
right-hand side to order
$g^n$, one obtains the distributional kernel of the $n$-th order obstruction of the
S-matrix, to be evaluated on $\chi^{\otimes n}$. It arises in terms of
$O_{U_{k_i}}$ acting on $L_k\pm K_k$, with
$k+\sum_{i=1}^rk_i=n$ (in particular only $k<n$ and $k_i<n$ appear). Remarkably, the odd part in $O_{U_{k_i}}$ is in
fact linear ($r=1$). At lowest orders: 
\bea{res}(L_2-K_2)-\pa U_2 &=& -i \, O_{U_1}(L_1+K_1), \\ \notag
(L_3-K_3)-\pa U_3 &=&\sfrac12 \, O_{U_1}^2(L_1-K_1) -\sfrac{3i}2\,
\big(O_{U_1}(L_2+K_2)+ O_{U_2}(L_1+K_1)\big),\\ \notag
(L_4-K_4)-\pa U_4 &=&
O_{U_1}^2(L_2-K_2) + (O_{U_1}O_{U_2}+O_{U_2}O_{U_1})(L_1-K_1) \\
\notag && -2i\,O_{U_1}(L_3+K_3)-3i\,O_{U_2}(L_2+K_2)-2i\,O_{U_3}(L_1+K_1) . \notag\eea
The symmetric re-installment of arguments $x_1,\dots,x_n$ and implicit delta
functions that are suppressed in the present simplified notation, is
unequivocal.  One must then evaluate the right-hand sides and resolve the
result by $L_n$ and $K_n$ along with $U^\mu_n$ on the left-hand side.
\\[1mm] 
(iii) \rref{r:pcb} applies similarly: When both $L_1$ and
$K_1$ satisfy the power-counting bound for renormalizability, then
$L_n=K_n=0$ for $n>2$. The third example in \aref{a:Exa} is an
instance where $K_1$ is not
renormalizable, and $L_n$ terminates but $K_n$ does not.
\\[1mm]
(iv) The right-hand sides of \eref{res} are the {\em modified} $n$-th order
obstructions of the S-matrix remaining {\em after} the passage from
$V$ to $U$, departing from \eref{O2defLV} (for $n=2$)  as
explained in \sref{s:main}. In analogy with \eref{LQn}, they
  may be
called ``LU-obstructions'' $i^{n+1}O^{(n)}_{\rm LU}$.
\\[1mm]
(v) When comparing \eref{res} with the second and third-order
formulas given in \cite{AHM}, one should observe, apart from the
reparametrization from $V$ to $U$, that in
\cite{AHM} the present $V(x;x')$
was expanded into delta functions and their derivatives  (where special feature of the model were exploited).%
\footnote{Specifically, \cite{AHM} included besides ``$V_2$'' also
``$W$-terms'' $\pa\pa'(W_2\cdot\delta_{xx'})$. If we  rename $V_2$
and $W_2$ from \cite{AHM} as $\wh V_2$ and $\wh W_2$ to prevent
confusion with the present $V_2$, then the relation $U(V)$ at second order is
$$U^\mu_2(x;x') = 2\wh V^\mu_2(x)\cdot \delta_{xx'} -
\pa'_\nu\big(\wh W_2^{\mu\nu}(x)\cdot\delta_{xx'}\big) + iO_{V_1(x')}(V^\mu_1(x)).$$
With this, the second and third order resolutions \eref{res}, while
different in appearance, are
equivalent to \cite[Eq.~(2.21)]{AHM}.}
Moreover, a major distinction is that in \eref{res}, the obstruction maps act only on the 
interactions $L_k$ and $K_k$, thanks to the elimination of $\pa U$
in \cref{c:LV}. This further simplifies actual computations.
\\[1mm]
(vi) As emphasized in \sref{s:main}: We did not prove that the
obstructions on the right-hand sides of \eref{res}, when evaluated,
are of the form that {\em can} be 
resolved at all orders, with $L_n$ and $K_n$ enjoying the respective
specified ``salient properties'' of the model under consideration.
\eref{LpmK} only provides a simple formula for
the combinatorics on the right-hand side, needed to {\em display}
the obstructions at order $g^n$ in terms of obstruction maps, once the obstructions of orders
$k<n$ have been resolved. On the other hand, \eref{LKU} resp.\
\eref{KLU} {\em always} determine $L_n$ resp.\ $K_n$ as ``functions'' of
$U_n$ and $L_n$ resp.\ $K_n$, but possibly without the specified
features. Thus, by \pref{p:LKU}, \eref{LVprop} is  always true without special
features of either $K_n$ or $L_n$ (or both). In particular, in more
general contexts (see \sref{s:dress}), they may not
necessarily factor a total delta function, which is of course one of the
resolvability conditions necessary to interpret $K_n$ and $L_n$ in
\eref{res} as interaction densities.
\end{remk}

\section{Towards interacting quantum fields}
\label{s:dress}
Interacting fields as operator-valued distributions are defined perturbatively by Bogoliubov's formula
\bea{Bog}\Phi\big\vert_{L}(f) = \int dx\, f(x)\Phi\big\vert_{L}(x):=-i\big(Te^{i\int dx\,
  \chi(x) L(x)}\big)^*\pa_s\big(Te^{i\int dx\, (\chi(x) L(x)+
  sf(x)\Phi(x)}\big)\big\vert_{s=0}
\eea
in the adiabatic limit $\chi(x)\to 1$. 

\eref{Bog} expands the interacting field into retarded integrals over
Wick products of free fields. In local QFT, these expansions are not
relatively local (in the sense of commutation
relations) to the free fields, but they are known to be relatively local
among each other.

The same formula is employed in sQFT, but if $L$ is string-localized,
it does not allow
to assess the localization properties of the interacting fields
$\Phi\vert_L$ relative to each other. This critical issue can be
settled with the help of the ``dressed field''.

\subsection{The dressed field}

For a free Wick polynomial $\Phi(x)$, we define the dressed field
$\Phi_{[g\chi]}(x)$ such that \eref{magic1} holds for
the interacting fields with a cutoff function:
\bea{magic}
\Phi\big\vert_{L[\chi]} (f)= \Phi_{[g\chi]}\big\vert_{K[\chi]}(f).
\eea
The task is to show that such a field exists, and to give a formula
for it. Setting $\chi=1$, gives \eref{magic1}. As we do not address
renormalization \cite{EG} in this paper, we claim and prove \eref{magic} only at
tree level.

\begin{prop}\label{p:dress} (Dressed field). With a cutoff function $\chi$, the dressed
field is the power series in $g\chi$
\bea{Phig}\Phi_{[g\chi]}(f) = \exp (iO_{U}[\chi])(\Phi(f)),
\eea
with the obstruction map $O_U$ as in \sref{s:LVall} understood, at
each order $n$, as a distribution in $x_1,\dots,x_n$.
\end{prop}

{\em Proof:} 
Assume that \eref{LVprop} holds without the insertion. Then, by \eref{Bog}, we have to
determine the dressed field such that
\bea{pasT} \pa_s Te^{i(K[\chi]+s\Phi_{[g\chi]}(f))}\big\vert_{s=0} = \pa_s
Te^{i(L+V^\mu\circ\pa_\mu)[\chi]+s\Phi(f))}\big\vert_{s=0}
\eea
holds. Let $K_s= K[\chi]+ s \Phi_{[g\chi]}(f)$. By \pref{p:LKU}, \eref{LVprop}
holds with $K$ and $L$ replaced by $K_s$ and $L_s$ where $L_s$ is
computed with \eref{LKU}:
$$ L_s = \exp( -iO_U)(K_s) + F(-iO_U)(\pa U)= L + s\exp(-iO_{U}[\chi])(\Phi_{[g\chi]}(f)).$$
This argument is legitimate because no specific properties of the
power series $K$ and $L$ were assumed in the proof of \pref{p:LKU}, see \rref{r:R}.(vi).

Thus, \eref{LKU} with $K$ and $L$ replaced by $K_s$ and $L_s$ gives
\eref{pasT} (even without the derivative w.r.t.\ $s$) provided
$$\exp(-iO_U[\chi])(\Phi_{[g\chi]}(f))= \Phi(f).$$
Applying $\exp(iO_U[\chi])$, we get \eref{Phig}. \qed

In local QFT, because the result of obstruction maps $O_{Y(x)}(X(x'))$
are always localized at $x=x'$, the dressed field $\Phi_{[g]}(x)$ is a power series of
Wick polynomials localized at the same point $x$.

In sQFT, due to the obstruction
maps involving string-localized fields $U$, the dressed field $\Phi_{[g\chi]}(x)$ will be a
perturbative series in Wick products of 
$g^n\Phi_{[n]}(x; x_1,\dots,x_n)$ evaluated on $\chi^{\otimes n}$, where
$\Phi_{[n]}(x; x_1,\dots,x_n)$ are supported at $x_i$ lying on the
string or cone emanating from $x$ or from $x_j$. In particular, if the
cone is convex, then $\Phi_{[n]}$ are string-localized. 

In the adiabatic limit $\chi\to1$, one has to integrate over all $x_i$. The
dressed field then becomes a power series in
$g$ whose coefficients may be string-integrals over Wick products of
(string-integrals over Wick products of \dots) string-localized free fields.

For  nontrivial examples of dressed fields in various models, see the appendix.

As a corollary to \pref{p:dress}, we get
\begin{coro}\label{c:autom} The ``dressing transformation''
  $\Phi\mapsto\Phi_{[g]}$ is an automorphism of the algebra $W[[g]]$ of power
  series in $g$ whose coefficients are Wick polynomials.
\end{coro}
{\em Proof:} The derivation property \eref{wickderiv} of the obstruction map $U$ implies
\bea{OU12}O_U^r(\colon \Phi_1\Phi_2\colon) = \sum_{r_1+r_2=r}
\sfrac{r!}{r_1!r_2!} \colon O^{r_1}_U(\Phi_1)O^{r_2}_U(\Phi_2)\colon.
\eea
\eref{OU12} implies the homomorphism property $\exp (iO_U)(\colon\Phi_1\Phi_2\colon) = \colon\exp
(iO_U)(\Phi_1)\exp (iO_U)(\Phi_2)\colon$ of its exponential. 
The inverse is $\exp (-iO_U)$. \qed

The following proposition shows that the dressing transformation
$e^{iO_U}(\cdot)$ also arises as an {\em interaction} (through Bogoliubov's
formula) with a suitable interaction $L_{\rm dress}$. 
\begin{prop}\label{p:Ldress} With the interaction density
  \bea{Ldress} L_{\rm dress} := F(-iO_U)(\pa U) = L - e^{-iO_U}(K),
  \eea
it holds
\bea{Lmagic}
\Phi_{[g]} = \Phi\vert_{L_{\rm dress}}.
  \eea
 \end{prop}
{\em Proof:} The equality of the two expressions in \eref{Ldress}
is \eref{LKU}. Now, let us regard $L=L(U,K)$ as a function of $U$ and
$K$, given by \eref{LKU}, which implies $\Phi\vert_L=
\Phi_{[g]}\vert_K$. Now, for $K$ and $U$ given, define $L_{\rm
  dress}:=L(U,0)$. Thus, it holds $\Phi\vert_{L_{\rm dress}}=
\Phi_{[g]}\vert_0$. This is \eref{Lmagic}.
\qed
\begin{remk}\label{r:Ldress}\rm
(i) In general, interactions do not produce automorphisms of the
  Wick algebra of free fields. This is the case here because at first
  order, $L_{\rm dress}$ is a total derivative: $(L_{\rm dress})_1=\pa
  U_1$, and the higher orders are  essentially ``quantum
    corrections'', necessary to cancel obstructions in order to get the 
  trivial $S$-matrix for an $L$-$V$ pair with   $K_{\rm dress}=0$.
\\[1mm]
(ii) The formal result of \pref{p:Ldress} is presumably not particularly useful. E.g., in the third example of
\aref{a:Exa}, where the issue is renormalizability, it turns out that
$L_{\rm dress}$ is a non-terminating series violating the power-counting bound. 
\end{remk}

Of prominent interest is the case when $\Phi_{[g]}(x)$ is a power
series of Wick products of {\em local} free fields, i.e., $\Phi_{[g]}$ is local relative to other free fields. This will not be the case in general.

Free fields $\Phi$ for which $\Phi_{[g]}$ is a local free
field (order by order), were called ``seeds for local interacting
fields'' in
\cite[Sect.~3.6]{AHM}.  Namely, because Bogoliubov's formula with a local
interaction preserves the relative localization, the corresponding interacting fields in
the adiabatic limit $\Phi\big\vert_L = \Phi_{[g]}\big\vert_K$ are local
relative to each other and to other interacting fields.

The question was raised in \cite{AHM} whether the seeds form
an algebra under the Wick product.
Thanks to \pref{p:dress}, the answer is affirmative.  

\begin{prop}\label{p:alg} (Algebra property of seeds). The seeds of local interacting fields form an algebra
under the Wick product. 
\end{prop}
{\em Proof:} $\Phi_i$ ($i=1,2$) are seeds if and only if all
$\Phi_{i,[n]}$ are local. By \pref{p:dress}, this is the case if and only if all $U^{r}_U(\Phi_i)$ are
local. By \eref{OU12}, it follows that $\colon \Phi_1\Phi_2\colon$ is a seed if $\Phi_i$ are
seeds. \qed

\section{Conclusion and outlook}
\label{Out}

We have reported substantial progress on the combinatorial structure of the
recursive scheme to control the effect of ``adding derivatives to the
interaction'' (and eliminate it by resolving obstructions).

Our main motivation is string-localized QFT, which allows to
reformulate (and in fact predict) Standard Model interactions in an
autonomous way (intrinsically quantum, referring neither to canonical quantization nor 
gauge or BRST invariance) in the ``LQ setting'' (\sref{s:LQall}) where the
string-variation at first order in the coupling constant is given by
derivatives. In the ``LV setting'', sQFT also allows to show equivalence
with gauge theory approaches (\sref{s:LVall}) where interactions that differ by
a total derivative at first order give rise to the same S-matrix and
interacting fields.

Yet, the results found are independent of the special application to
sQFT, and may as well be used to compare, say, different local methods to
deal with interactions of massive vector bosons. (An example is given in \cite[Sect.~4.5]{AHM}.)

Renormalizable interactions are Wick polynomials of at most
quartic degree. This means that the corresponding power series in $g$
terminate at second order, see \rref{r:pcb}. Yet, the validity of our
analysis at all orders is relevant for several reasons:
\\[1mm]
(i) Charged dressed fields are in general non-terminating power
series, which is legitimate because they are not observables. Yet they
are important elements of interacting QFT models.
\\[1mm]
(ii) Quantum field theories beyond the power-counting bound are within
the scope of our method, notably perturbative GR \cite{Dü2,GGR}. One may even
speculate that the restrictions imposed by the resolvability of
obstructions at all orders might help to fix infinitely many
(otherwise free) renormalization constants at all orders.

An interesting but open question is how to pass from LV to LQ. At first order this is
trivial because each $L$-$V$ pair with $\ddelta K_1=0$ gives rise to
an $L$-$Q$ pair by $Q_1^\mu = \ddelta V_1^\mu$. At higher orders the
connection is less clear because $\ddelta$ acts both on $L$ and on $V$
(or $U$). The expectation remains that ``LQ is some kind of
infinitesimal version of LV''.

\appendix

\section{Examples in sQFT}
\label{a:Exa}

Many applications of sQFT to the Standard Model have been elaborated,
at different level of systematization, in
\cite{GGM,GGR,GMV,GRV,Infra,AHM,nab}. For a review, see
\cite{Aut}. Here, we present just a selection of issues concerning the
dressed field.

\paragraph{1. sQED.}
We illustrate the dressed field by the example of 
dressed Dirac fields in string-localized QED. Besides the $L$-$Q$ pair
\eref{QEDLQ} on the Hilbert space, there is also an $L$-$V$ pair
embedded into the indefinite space (Krein space) of gauge theory. It looks just like
\eref{ProcaLV} but with a massless field $\phi(x,c)$ (the
string-integrated gauge potential that lives in the Krein space).

All relevant obstruction maps vanish on neutral
fields, so that there are no induced interactions (\sref{s:LVall} is
void) for sQED \cite[Sect.~3.1]{Infra}. But the violation
of the charged Ward identity mentioned in \sref{s:glimpse} implies that
$O_{U_1}(\psi)= \phi(c)\cdot\psi$. Pushing to higher orders, one gets
the dressed Dirac field \cite{Infra}
\bea{ddir} \psi_{[g]}(x)= \wick{e^{ig\phi(x,c)}}\cdot \psi(x).\eea
The exponential is a Wick power series
in the string-integrated free gauge potential. It can be viewed as a smeared abelian Wilson
operator%
\footnote{Quantities of the form \eref{ddir} were 
considered by many authors \cite{Jor,Dir,Man} as {\em
  classical} expressions with the motive to quantize only gauge
invariant quantities. Steinmann's idea \cite{Stein} to
formulate perturbation theory as a perturbation of a free {\em
  quantum} field \eref{ddir} is 
very close to ours in \eref{magic}, with the distinction that
Steinmann {\em chose} \eref{ddir}, while sQFT {\em derives} it.},
which is itself a string-localized field. In fact, due to the infrared
divergence of the field $\phi$, {\em only} its exponential can be
defined. In \eref{ddir}, it turns $\psi_{[g]}$ into an ``infrafield'', creating states
on which the mass operator has no point spectrum. This feature
reflects the ``photon cloud'' accompanying a charged particle.

The string-localization of the dressed, hence of the interacting Dirac
field is also crucial to resolve the conflict between Gauß' Law and
Locality \cite{FPS}: the total charge operator is a surface integral localized at spacelike
infinity, hence commutes with all local fields. Thus, it cannot
``see'' the charge of a local charged field.

\paragraph{2. sQCD.}
Similar as sQED, but with obstruction maps giving nontrivial results on the
nonabelian currents, and the Lie-algebra valued field $\phi(c)$ is not ``inert''
in obstruction maps involving the nonabelian $F_{\mu\nu}$.

Consequently, there are induced interactions (corresponding to the quartic gauge
interaction), and the  dressed quark triplet field in string-localized QCD is more 
complicated \cite{nab}. We have 
computed it until third order, finding that the exponential in \eref{ddir} is
replaced by a power series in string-integrals over Wick products of
string-integrals over Wick products of string-localized
Lie-algebra valued free fields. The nested structure of these
integrals precisely amounts to
path-ordering in the case when the string is just a line from $x$ to
$\infty$. In the smeared case, the exponential still enjoys properties of a ``smeared
Wilson operator''%
\footnote{We are not aware of an independent notion of ``path-ordering'' for exponentials of smeared line integrals.};
in particular, the dressed field is invariant when classical gauge transformations that are trivial
at infinity act on the free quantum gauge potential and Dirac field it
is made of.

\paragraph{3. Abelian Higgs model.}
Another example from sQFT that can
be solved in closed form illustrates the nontriviality of the dressed
field in general.

The field content of the abelian Higgs model is a Proca field $B_\mu$
of mass $m$ and a scalar Higgs field $H$ of mass $m_H$. Because the Proca field has UV dimension
2, the cubic local
interaction density
\bea{K1} K_1(x)=m\big(B_\mu B^\mu H + aH^3\big)(x)
\eea
is non-renormalizable by power counting. With the help of
\eref{ABphi}, one gets
an $L$-$V$ pair
\bea{L1} L_1(x,c) &=& m\big(A_\mu(c) (B^\mu H + \phi(c) \pa^\mu
H)-\sfrac12m_H^2\phi(c)^2 H +aH^3\big)(x) = K_1(x) + \pa_\mu V_1^\mu(x,c),
\notag\\ V_1^\mu(x,c) &=& m\big(B^\mu\phi(c) H+ \sfrac12 \phi(c)^2 \pa^\mu H\big)(x),
\eea
where $L_1$ is renormalizable by power counting. The parameter $a$ is
undetermined at this stage. One may add another $L$-$V$ pair for the
coupling to a Dirac field, which we ignore for the
sake of the example.%
\footnote{Minimal interactions of several Proca fields via non-abelian currents are
  separately inconsistent and are made consistent by
  adding self-interactions {\em and} \eref{L1}, see \cite{nab}. This
  was mentioned in \sref{s:Intro}.}
For more details, see \cite{AHM}. 

The self-propagators of the fields $B_\mu$ and $\pa_\mu H$ admit 
delta-function renormalizations with arbitrary coefficients $c_B$ and
$c_H$. It turns out that the higher obstructions are 
resolvable only with the choice $c_B=c_H=-1$. With these values, all
relevant obstructions are multiples of $i\delta(x-x')$ which is the
reason why  the
model is particularly simple. We computed $L_n$, $K_n$, and
$U_n$ until order $g^5$. At order 2,
\bea{KL2} K_2&=& -3B^\mu B_\mu H^2 +bH^4,  \notag \\
L_2 &=& 
-\sfrac14m^2m_H^2\phi^4-\sfrac12 m_H^2\phi^2H^2+bH^4 
\eea
with another free parameter $b$. At order 3, the values of the
Higgs self-coupling constants $a$ and $b$ are fixed. Then all
obstructions are resolvable with all $L_n=0$ for $n>2$, so that the
power-counting bound is respected. $K_n$ for $n>2$ are the coefficients of the
power series $-\frac12B_\mu B^\mu (1+\frac gm H)^{-2}$, which are
increasingly non-renormalizable. $U_n^\mu$ are polynomials in $\phi$ and $H$
multiplying $B^\mu$ and $\pa^\mu H$.

The dressed fields until order $g^5$ allow to guess the
closed formulas
\bea{AHMg}
A^\mu_{[g]}=A^\mu, \quad mB^\mu_{[g]}=
\cos(g\phi)\cdot\frac{mB^\mu}{1+\frac gm H} - \sin(g\phi)\cdot\pa^\mu
H,\quad g\phi_{[g]} = \sin(g\phi)\cdot (1+\frac gm H)\notag\\
\frac gm H_{[g]} = \cos(g\phi) \cdot (1+\frac gm H)-1, \quad
(\pa^\mu H)_{[g]}=\sin(g\phi)\cdot\frac{mB^\mu}{1+\frac gm
  H}+\cos(g\phi)\cdot \pa^\mu H.\quad
\eea
Assuming these formula to be exact at all orders, one can determine
$U_n$ also beyond $n=5$ with the help of \eref{Phig},  without having
to compute and resolve obstructions. Having confirmed this ``prediction'' also at order $6$, we
refrained from going further.

We don't have a candidate closed formula for $U$ at all orders, and
neither for the ``interpolating'' densities $\ell(t)$ of
\pref{p:LVt}. But it is clear without any computation that $\ell(t)$
can be renormalizable only at $t=1$. Namely, each $\ell_n(t)$ is a
polynomial in $t$ (of degree $n$), interpolating $\ell_n(0)=K_n$
(non-renormalizable) with $\ell_n(1)=L_n$ (renormalizable, in fact
$L_{n\geq3}=0$).  The fact that all $\ell_{n\geq3}(t)$ have
simultaneous zeroes at $t=1$ is highly nontrivial, but  
indispensable for resolvability at all orders. It is not explained by 
\rref{r:R}.(iii).

The transformations $\Phi\mapsto\Phi_{[g]}$ in \eref{AHMg}
are invertible (in terms of algebraic functions), confirming \pref{c:autom}. They do
not form a one-parameter group w.r.t.\ $g$ (which is not expected),
but they are the particularly nice values at $t=1$ of the (unknow to us) one-parameter transformation group $e^{itO_U}(\cdot)$.

This paper is not the place for physical discussions of
\eref{AHMg}. But it is clear that complex 
combinations of fields have dressing transformations involving the
string-localized exponentials $e^{\pm ig\phi(c)}$, generalizing
\eref{ddir}. The same factor will also arise as a dressing of a
  Dirac field minimally coupled to the string-localized Proca field.
Unlike the massless case, $\phi(c)$ is IR regular and has UV 
dimension 1. It would be interesting to explore whether its string-localization
helps to cure the long-known problems with exponentials of free local massive
scalar fields $\varphi$ \cite{Ja}, namely, the non-existence of test
  functions $f(x)$ that would turn $e^{ig\varphi(x)}$ into a local
  operator. Allowing string-localization because $e^{ig\phi(x,c)}$
is string-localized anyway, might relax the problem.

\paragraph{Acknowledgment.} The author is indebted to the
  anonymous referee who incited substantial improvements of a previous
version.

\end{document}